\RequirePackage{fix-cm}
\documentclass[twocolumn]{svjour3}          
\smartqed  
\usepackage{graphicx}
\usepackage{xcolor}
\usepackage{amsmath}
\usepackage{amssymb}
\usepackage{hyperref}

\newcommand{\cmmnt}[1]{\ignorespaces}
\usepackage{ulem}

\begin{document}

\title{Hubbard-like Interactions and Emergent Dynamical Regimes Between Modulational Instability and Self-Trapping}


\subtitle{}


\author{L. J. R. Bezerra Jr. and W. S. Dias*}

\institute{L. J. R. Bezerra Jr. \at Instituto de F\'{\i}sica, Universidade Federal de Alagoas, 57072-970 Macei\'o, AL, Brazil \email{luciano.bezerra@fis.ufal.br} \\ W. S. Dias \at Instituto de F\'{\i}sica, Universidade Federal de Alagoas, 57072-970 Macei\'o, AL, Brazil \email{wandearley@fis.ufal.br} 
}



\date{Received: date / Accepted: date}

\maketitle

\begin{abstract}
We investigate the modulational instability of uniform wave packets governed by a discrete third-order nonlinear Schrödinger equation in finite square lattices, modeling light propagation in two-dimensional nonlinear waveguide arrays. We analyze how initially stable uniform distributions evolve into self-trapped (localized) regimes and the influence of a refractive index detuning selectively applied along the diagonal waveguides on this transition. This detuning effectively emulates the effect of on-site Hubbard-like interactions $U$ in photonic analogs of interacting particles in a one-dimensional lattice. While for $U = 0$ the system exhibits the known direct transition from stable uniform states to asymptotically localized profiles, we show that $U > 0$ induces an emergence of intermediate dynamical regimes. These regimes include coherent breathing modes that can be either confined along diagonal or off-diagonal waveguides, as well as chaotic-like propagation patterns. At higher nonlinearities, we identify distinct self-trapped regimes characterized by diagonal or off-diagonal localized modes, depending on the strength of $U$. The critical nonlinear strengths separating the existing regimes are shown in the phase diagram, underscoring the competing trends imposed by the Hubbard-like interaction on the optical field. 
\end{abstract}

\section{introduction}
\label{introduction}

Nonlinear effects have attracted significant attention in various branches of science for their role in generating complex and unique behaviors. In solid-state physics, these effects are inherent to understanding interactions between elementary excitations and lattice vibrations, giving rise to phenomena such as soliton formations and breather modes~\cite{HOLSTEIN1959325,PhysRevB.47.15330,PhysRevB.51.15038,MORAIS2020166798}. Nonlinearities resulting from changes in the refractive index of the medium lead to the emergence of striking features in optics, such as solitons and self-trapping~\cite{HENNIG1999333,RevModPhys.77.633,LEDERER20081}. In Bose-Einstein condensates, studying nonlinear interactions has revealed distinctive behaviors, including the emergence of breathers and self-trapped formations~\cite{RevModPhys.71.463,RevModPhys.78.179,PhysRevLett.92.040401,Strecker2002}. Additionally, rogue waves and distinct soliton-like formations have been reported in quantum information systems~\cite{PhysRevA.75.062333,PhysRevA.101.023802,Angles-Castillo_2024,Maeda_2019,PhysRevA.106.012414}. The scientific and technological progress stemming from these studies underscores the need for a broader understanding of the mechanisms underlying these phenomena. Such a scenario becomes essential for propelling technological advancements, optimizing material properties, and enhancing control over physical protocols within diverse scientific domains.


Modulational instability,  a phenomenon which addresses the stability of constant-amplitude waves under perturbations in nonlinear dispersive media~\cite{ZAKHAROV2009540,Hasegawa1990}, is recognized as a precursor to various complex regimes~\cite{PhysRevA.65.021602,Tchepemen2023,Tang2017,DIAS2019121909,Bezerra2023,Belic2022}. In Bose-Einstein condensates within one-dimensional optical lattices, it plays a key role in generating bright and dark soliton-like wave functions~\cite{PhysRevA.65.021602}. Properly tuning higher-order nonlinearities and nonlocal response functions in composite systems leads to the emergence of Akhmediev breathers~\cite{Tchepemen2023}. Nonlinear discrete lattices including next-nearest-neighbor couplings reveal the emergence of breather solutions with both dark and bright profiles, depending on the coupling strength~\cite{Tang2017}. Long-range couplings exhibiting a power-law can induce even more intricate dynamics, including breathing and chaotic-like patterns, which act as intermediate states between stable uniform waves and localized solutions~\cite{DIAS2019121909}. Recent findings have pointed to modulational instability as a potential mechanism for the formation of rogue waves and extreme events in nonlinear discrete lattices~\cite{Bezerra2023,Belic2022}. 

Beyond the characterization of distinct dynamical regimes, modulational instability has also been explored for potential applications. Its inherent sensitivity to perturbations has been proposed as a tool for probing band topology in photonic lattices~\cite{PhysRevLett.126.073901} and for analyzing the response of quantum walks to instantaneous decoherence~\cite{PhysRevA.103.042213}. In dusty plasma systems, modulational instability has been considered as a key mechanism behind freak oscillations, where nonlinear coupling and dispersion give rise to transient, large-amplitude localized structures from uniform states~\cite{PhysRevE.95.053207}.

Although modulational instability sets the critical conditions for the breakdown of uniform solutions, the dynamical pathways connecting stable uniform states to localized regimes remain less explored. The emergence of intermediate dynamical regimes, such as regular or chaotic-like breathing modes, is strongly influenced by both the system dimensionality and specific interaction mechanisms. In some systems, this transition may occur directly, with no intermediate dynamical regimes, as reported for square lattices by Chaves \textit{et al.} ~\cite {Chaves2015}. More generally, both the thresholds and the presence of intermediate dynamical behaviors are sensitive to the system’s effective dimensionality~\cite{DIAS2019121909}. For spectral dimensions $d_{\text{eff}} < 2$, the threshold instability decreases with increasing system size, suggesting that uniform regimes may become unstable under arbitrarily weak nonlinearities in the thermodynamic limit~\cite{DIAS2019121909}. Moreover, additional ingredients such as saturable nonlinearity can enhance the stability of uniform wave solutions, postpone the onset of soliton-like structures, suppress the regular breathing regime, and enlarge the chaotic-like regime where rogue waves become more likely to occur~\cite{Bezerra2023}.

We observe that two-dimensional lattice models have recently gained prominence as versatile platforms for simulating quantum systems of interacting particles, particularly in photonic settings~\cite{Schwartz2007,PhysRevA.78.033834,PhysRevB.103.214311,PhysRevLett.123.043201}. Motivated by this dual perspective, which involves the established direct transition in nonlinear lattices and the potential of photonic platforms for quantum simulation, we study here how initially stable uniform distributions evolve into self-trapped regimes in a two-dimensional discrete system. Our focus lies on the role of a selective refractive index detuning applied along the diagonal waveguides, which effectively emulates an on-site Hubbard-like interaction in photonic analogs of two interacting particles in a one-dimensional lattice~\cite{Longhi:11,PhysRevA.89.023823,Corrielli2013,PhysRevB.76.155124}. By varying both the nonlinear strength and the amplitude of the Hubbard-like interaction, we uncover the emergence of intermediate dynamical regimes, along with the existence of distinct self-trapped patterns. The critical nonlinear thresholds separating these regimes are summarized in the phase diagram, underlining the competing effects of the Hubbard-like interaction on the system's dynamical evolution.

\section{model}
\label{model}

The problem involves analyzing a set of discrete nonlinear Schrödinger equations with $\hbar = 1$, written as
\begin{eqnarray}
i\dfrac{dc_{n,m}}{dt} &=& (\beta_{n} + \beta_{m})c_{n,m} + \kappa(c_{n-1,m}+c_{n+1,m}+ c_{n,m-1} \nonumber\\
&+&c_{n,m+1})+\chi|c_{n,m}|^2c_{n,m}+ U\delta_{n,m}c_{n,m}
\end{eqnarray}
These equations are analogous to the coupled-mode equations describing light propagation in a symmetric square 2D waveguide array, where $c_{n,m}$ denotes the amplitude in the individual waveguides. $\beta_{n}$ and $\beta_{m}$ represent the propagation constants along the respective lattice directions, $\kappa$ quantifies the evanescent coupling between adjacent waveguides, and $\chi$ characterizes the third-order Kerr nonlinearity of the waveguide material~\cite{Christodoulides:88,Yaron2003}. The parameter $t$ corresponds to the longitudinal propagation coordinate and plays the role of an effective time within the paraxial approximation. For $U=0$, the analysis of modulational instability reports a direct transition from a regime characterized by stable uniform to one characterized by asymptotically localized solutions~\cite{Chaves2015}. Here, we investigate the influence of a selective refractive index detuning or waveguide width modulation applied along the diagonal waveguides ($n = m$), which effectively emulates an on-site Hubbard-like interaction $U$ in photonic analogs of two interacting particles in a one-dimensional lattice~\cite{Longhi:11,PhysRevA.89.023823,Corrielli2013,PhysRevB.76.155124}.


In the following analysis, we assume the absence of disorder, thereby avoiding related features associated with Anderson localization~\cite{PhysRev.109.1492}. With a uniform coupling, we set the field propagation constant ($\beta_{n(m)}=0$) without loss of generality. Furthermore, all quantities are expressed in dimensionless units by setting $\kappa=1$, which implies that $\chi$ and $U$ are measured in units of $\kappa$.

As part of our analysis of modulational instability, we consider an initial wave packet composed of a uniform background amplitude $c_0$ superimposed with a weak random perturbation of strength $\epsilon = 10^{-3}c_0$. We study square lattices of size $L \times L$, where $L$ is the number of sites (waveguides) along each direction, resulting in a total of $N = L^2$ sites. The uniform amplitude is set to $c_0 = 1/\sqrt{N}$, and the initial amplitudes at each site are randomly distributed within the interval $[c_0 - \epsilon, c_0 + \epsilon]$. After generating all site amplitudes, the wave function is normalized to ensure a unit total power.

The resulting set of differential equations is solved using a standard eighth-order Runge-Kutta method. A sufficiently small time step is chosen to ensure the conservation of wave function norm, adhering to the criterion ($|1 -  \sum_{n,m}|c_{n,m}(t)|^2 | \leq 10^{-14}$ ) throughout the entire time interval considered. Additionally, we assume periodic boundary conditions consistent with the methodology outlined in Ref.~\cite{Chaves2015,DIAS2019121909}


\section{Results and discussion}
\label{Results_and_discussion}  

\begin{figure*}[t]
    \centering
    \includegraphics[height=6.cm]{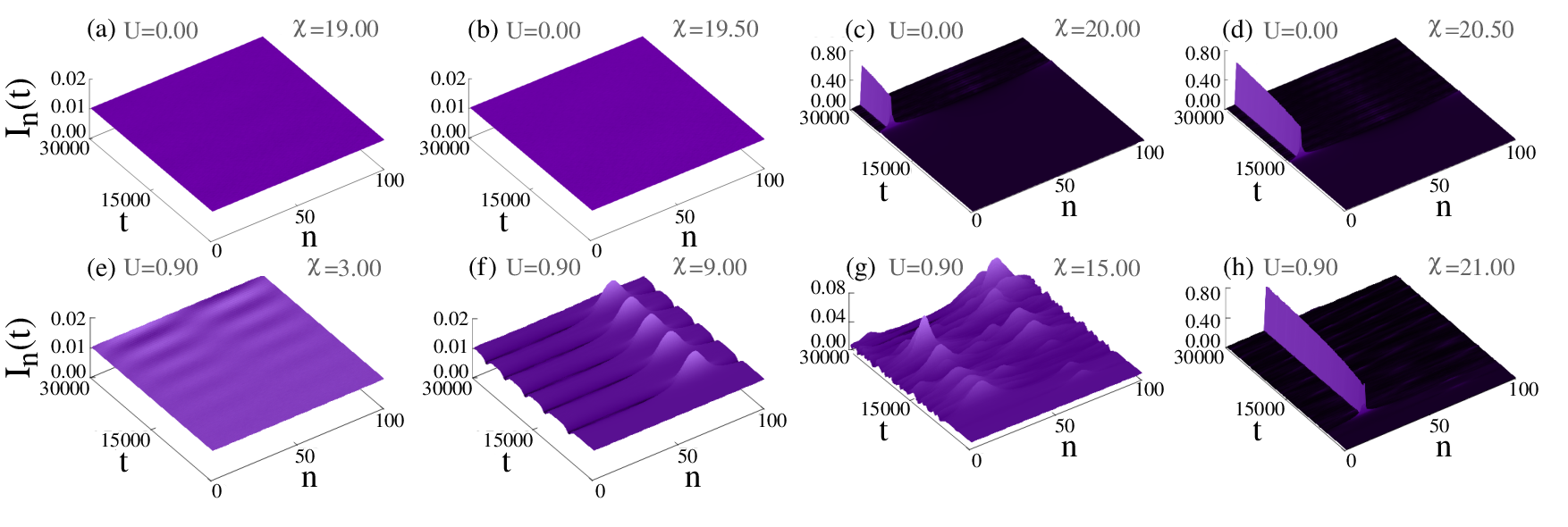}
    \caption{Time evolution of the marginal intensity profile $I_{n}(t)=|c_{n}(t)|^2$ along a single axis of a square waveguide lattice with $L = 100$, for representative nonlinear strengths $\chi$. In the absence of the on-site Hubbard-like interaction (panels a-d), a direct transition is observed from a regime of stable uniform light distribution to a stationary, strongly localized optical mode. This transition occurs around $\chi\approx 19.7$, consistent with previous reports. When a weak on-site Hubbard interaction is introduced (panels e-h), the direct transition is replaced by the emergence of intermediate breathing regimes. These results suggest that Hubbard-like interactions significantly narrow the range of $\chi$ over which the uniform optical distribution remains stable.} 
    \label{fig:1}
\end{figure*}
We start by monitoring the evolution of the light intensity distribution along one coordinate axis of the square waveguide array. Specifically, we compute the marginal intensity profile $I_{n}(t) = |c_{n}(t)|^2 = \sum_{m} |c_{n,m}(t)|^2$ for a square lattice with $L = 100$. Similar behavior is observed for $I_{m}(t)$, and is therefore omitted for brevity. In Fig.~\ref{fig:1}, we show the dynamics in systems governed by representative values of the nonlinear parameter $\chi$ and the Hubbard-like interaction $U$. In Figs.~\ref{fig:1}a-d, we present distinct dynamical regimes that emerge as the nonlinearity increases while keeping $U=0$. For small nonlinearities (Figs.~\ref{fig:1}a-b), the wave packet remains uniformly spread across the entire lattice, signaling the persistence of the initial uniform field distribution. A self-trapped optical field emerges by increasing the nonlinear parameter (see Figs.~\ref{fig:1}c-d). This transition occurs suddenly near $\chi \approx 19.7$, which is consistent with the findings in Ref.~\cite{Chaves2015}.
When we introduce a weak on-site Hubbard-like interaction ($U=0.90$), as shown in Figs.~\ref{fig:1}e-h, the direct transition from a stable uniform field distribution to a strongly localized pattern becomes mediated by intermediate breathing regimes. Moreover, the critical nonlinear strength required to preserve the stability of the initial state is substantially lowered. This behavior is attributed to the refractive index offset acting as an effective energy barrier for field confinement along the diagonal. Even weak detuning values such as $U = 0.90$ hinder the coupling between diagonal and neighboring waveguides, such as in the particle model, where even a weak Hubbard-like interaction inhibits transitions between doubly occupied (bound) and separated (unbound) states, thereby effectively decoupling the bound and unbound sectors of the Hilbert space. This constraint, combined with the third-order Kerr nonlinearity, amplifies local intensity fluctuations and contributes to the destabilization of uniform propagation, thereby increasing the system's susceptibility to modulational instability.


To better understand and characterize the emergence of signaled breather solutions, we analyze the time evolution of the directional participation function,
\begin{equation}
P_{i}(t) = \dfrac{1}{\sum_{i=1}^{L} |c_{i}(t)|^4}, \hspace{.5cm}i=n,m
\end{equation}
where $|c_{i}(t)|^2 = \sum_{j} |c_{i,j}(t)|^2$ denotes the optical intensity distribution along the $i$-th waveguide axis. The index $j$ refers to the transverse direction with respect to $i$, and the pair $(i,j)$ covers all sites in the two-dimensional waveguide array. This measure provides an estimate of the effective number of waveguides that are significantly excited along the $i$th direction, with $P_{i} \approx L$ in the case of a uniform distribution, while in the limit of complete localization, $ P_{i} \approx 1$.

\begin{figure}[t]
    \centering
    \includegraphics[height=6.7cm]{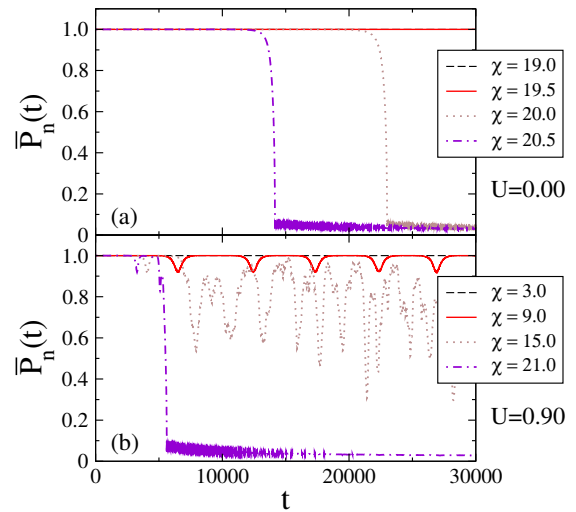}
    \caption{Time evolution of the normalized participation ratio for a single transverse direction [$\overline{P}_{n}(t) = P_{n}(t)/L$], analyzed for the same settings of $\chi$ and $U$ as in Fig.~\ref{fig:1}. In a stable uniform light distribution, the intensity profile remains fully extended over the lattice, yielding $\overline{P}_{n}(t) = 1$. Conversely, in a self-trapped regime, light becomes strongly localized, driving $\overline{P}_{n}(t) \approx 0$. The transition between these regimes is influenced by the presence of the diagonal Hubbard-like coupling, which induces intermediate breathing dynamics, both regular and chaotic-like, and modifies the critical nonlinear threshold required to achieve a stable uniform regime.} 
    \label{fig:2}
\end{figure}

In Fig.~\ref{fig:2} we display the normalized participation function $[\overline{P}_{i}(t) = P_{i}(t)/L]$ for $i=n$ for the same configurations of $\chi$ and $U$ presented in Fig.~\ref{fig:1}. Similar behavior is observed for $\overline{P}_{m}(t)$, and is therefore omitted for brevity. In the uniform regime, the optical field remains fully extended throughout the evolution, yielding ($\overline{P}_{n}(t) = 1$). Conversely,  in the self-trapped regime, the field becomes strongly localized, with ($\overline{P}_{n}(t) \approx 0$). In the absence of on-site Hubbard-like interaction ($U=0$; Fig.~\ref{fig:2}a), this transition occurs directly at $\chi \approx 19.7$, in full agreement with the previous report~\cite{Chaves2015}. When $U=0.90$ (see Fig.~\ref{fig:2}b), the participation function displays fluctuations in an intermediate regime of $\chi$. The behavior observed at $\chi=9.0$ is signaled by regular breaths, which periodically restore the fully extended profile. As  $\chi$ increases to $15.0$, the participation function develops irregular, aperiodic modulations with varying minima. These distinct breathing behaviors are consistent with previously reported regular and chaotic-like regimes~\cite{Bezerra2023,Chaves2015}.

\begin{figure}[tp]
    \centering
    \includegraphics[height=11.cm]{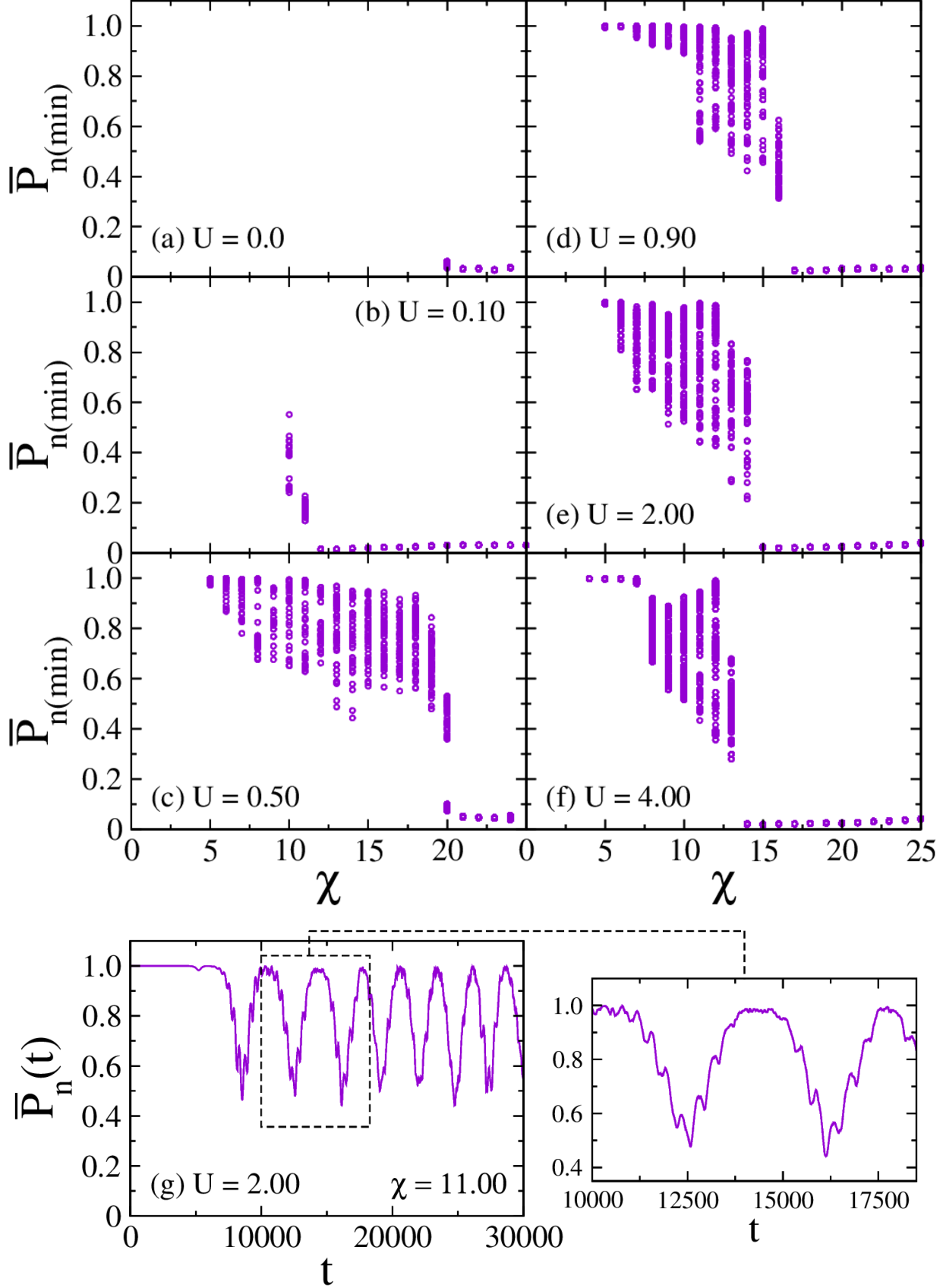}
    \caption{Minimum of the normalized participation function for a single transverse direction as a function of the nonlinear strength $\chi$ in waveguide arrays with representative Hubbard-like interactions ($U$). For $U=0$, the direct transition from stable uniform field distribution to self-trapped regimes with increasing $\chi$ is confirmed. In contrast, panels where $U>0$ signal this transition mediated by distinct breathing regimes, including regular and chaotic-like dynamics. A complementary analysis of the temporal evolution reveals the emergence of additional breathing modes to regular breathing (see Fig.~3g), posing challenges for mapping the chaotic-like regime.}
    \label{fig:3}
\end{figure}

Based on previous results, which show how the participation function characterizes the existing dynamical regimes, we computed the local minima of the normalized participation ratio for a wide set of nonlinear strengths. By accounting for a possible initial transient time, only minima occurring in the interval $10000 < t < 30000$ are written down. Fig.~\ref{fig:3}a-f shows plots of these minima as a function of $\chi$ for systems with representative Hubbard-like interaction $U$. For $U=0.0$ (see Fig.~\ref{fig:3}a), the minima are $\overline{P}_{{n}(min)} \approx 0$ and appear from the modulational instability threshold ($\chi \approx 19.7$), consistent with the known direct transition to the self-trapping regime~\cite{Chaves2015}. Intermediary values of $\overline{P}_{{n}(min)}$ signal the emergence of breathing regimes, which are observed in systems with finite Hubbard-like interaction (see Fig.~\ref{fig:3}b-f). These breathing modes arise even with a Hubbard-like interaction, as noted at $U=0.10$. Furthermore, we observe that the range of nonlinear strength ($\chi$) over which the initial uniform field distribution remains stable becomes narrower as $U$ increases. This scenario reflects how the diagonal modulation introduced by $U$ penalizes uniform propagation along the waveguides, amplifying local intensity variations and contributing to the destabilization of the initial homogeneous state.


Further analysis of Fig.~\ref{fig:3} reveals settings where the minima of the participation ratio are narrowly clustered, indicating a regular breathing dynamics. As the nonlinear strength $\chi$ increases, the system undergoes more intricate breathing modes. These regimes are characterized by a broader dispersion of participation minima, suggesting enhanced interference effects and the possible coexistence of multiple breathing modes with distinct frequencies. Importantly, not all such configurations correspond to chaotic-like behavior, as shown in Fig.~\ref{fig:2}b for $U = 0.9$ and $\chi = 15.0$. We have identified configurations in which the field profile exhibits regular breathing dynamics characterized by two dominant frequencies, repeatedly returning to the fully extended state throughout its evolution (see Fig.~\ref{fig:3}g).

\begin{figure}[t]
    \centering
    \includegraphics[height=8.9cm]{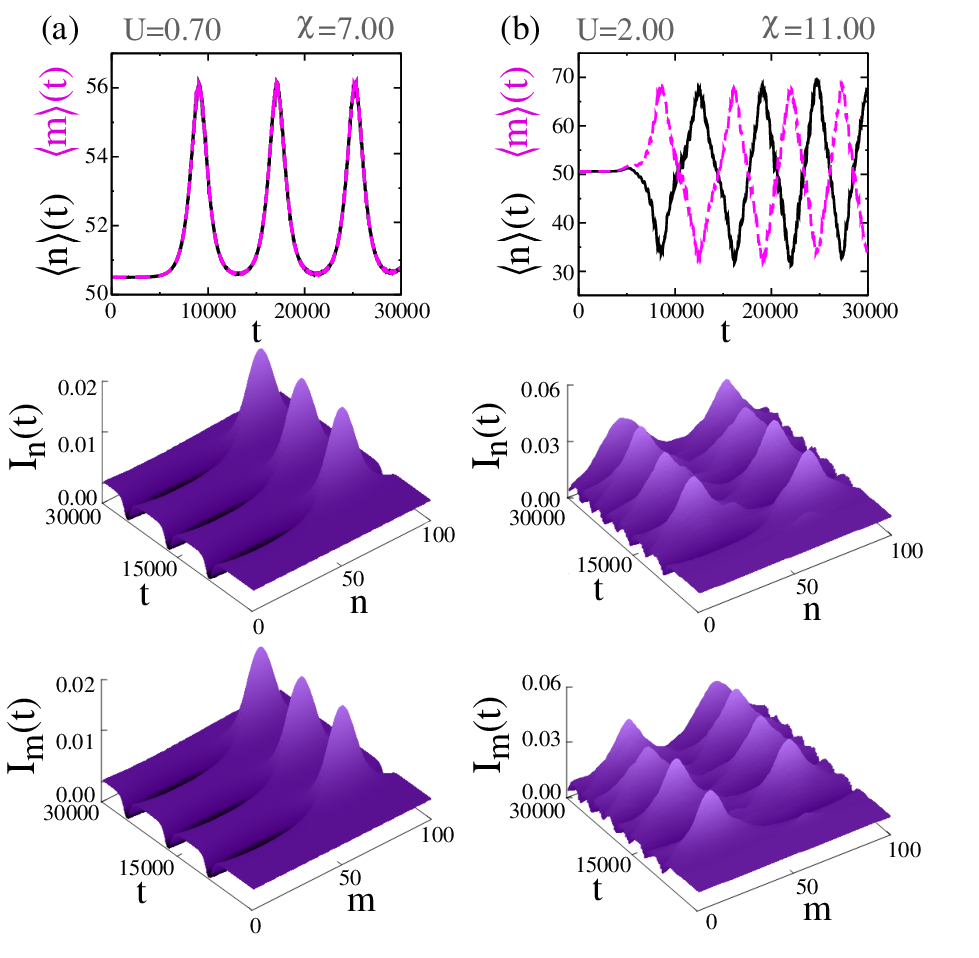}
    \caption{Time evolution of wave-packet centroids along both lattice directions and their corresponding intensity profiles, depicting two distinct breathing regimes arising from the competition between existing diagonal and off-diagonal modes. (a) Bound breathing regime: The field undergoes periodic amplitude modulation (breathing) while remaining localized at a diagonal waveguide ($n=m$). (b) Unbound breathing regime: The field exhibits a breathing pattern characterized by a periodic spatial shift between symmetric off-diagonal waveguides.}
    \label{fig:4}
\end{figure}

To better characterize the underlying mechanisms of this transitional regime between stable uniform and self-trapped (localized) regimes, we complement the analysis of the wave-packet profiles and their corresponding participation functions by examining the time evolution of the wave-packet centroid along each lattice direction, defined as
\begin{equation}
\langle n\rangle(t) = \sum_{n,m} n|c_{n, m}(t)|^2,  \hspace{0.3cm} \langle m\rangle(t) = \sum_{n,m} m|c_{n, m}(t)|^2.
\end{equation}
In Figs.~\ref{fig:4}a-b, we show the time evolution of the centroids $\langle n\rangle(t)$ and $\langle m\rangle(t)$, along with the corresponding wave-packet profiles projected along each lattice axis, for two representative configurations exhibiting breathing dynamics. In Fig.~\ref{fig:4}a, where $U = 0.7$ and $\chi = 7.0$, the intensity profiles show dynamically bound oscillations, predominantly concentrated along the diagonal elements of the array, consistent with coupled regular breathing modes. In contrast, the configuration with $U = 2.0$ and $\chi = 11.0$ (Fig.~\ref{fig:4}b) displays a coherent breathing pattern in which the optical field periodically oscillates across non-diagonal waveguides, effectively avoiding periodic intensity localization along the diagonal. The behavior suggests a suppression of diagonal mode propagation, analogous to the avoidance of double occupancy in repulsive interaction regimes. Such patterns reflect the underlying competition between existing states and their role in destabilizing uniform distributions.    

In the absence of nonlinearity and Hubbard-like interactions, the stationary solutions of the linear coupled-mode equations correspond to Bloch states of the periodic lattice. These solutions take the form $\psi_{n,m} \propto e^{i(k_1n+k_2m)}$, and their associated energy band covers the range [$-4\kappa, 4\kappa$], which effectively defines the band structure of the extended (unbound) modes in the waveguide array. Hubbard-like interaction is well-known for supporting the formation of bound states, defect modes that are localized at
the $n=m$ interface. Such a states covers the range $U \leq E \leq \sqrt{16\kappa^2 + U^2}$~\cite{Longhi:11,PhysRevB.76.155124}. Thus, the emergence of a weak interaction $U$ supports coherent dynamics in which the optical fields are predominantly confined along the diagonal ($n = m$), consistent with the breathing behavior observed in Fig.\ref{fig:4}a.   However, as the interaction strength $U$ increases, the subband associated with these (bound) modes becomes progressively detached from the main (unbound) band. This spectral separation introduces a growing phase mismatch and acts as an effective barrier for diagonal coupling, thereby suppressing the coherence of diagonal modes. Consequently, the optical field redistributes along off-diagonal paths, in agreement with the dynamics shown in Fig.\ref{fig:4}b.

\begin{figure}[t]
    \centering
    \includegraphics[height=8.9cm]{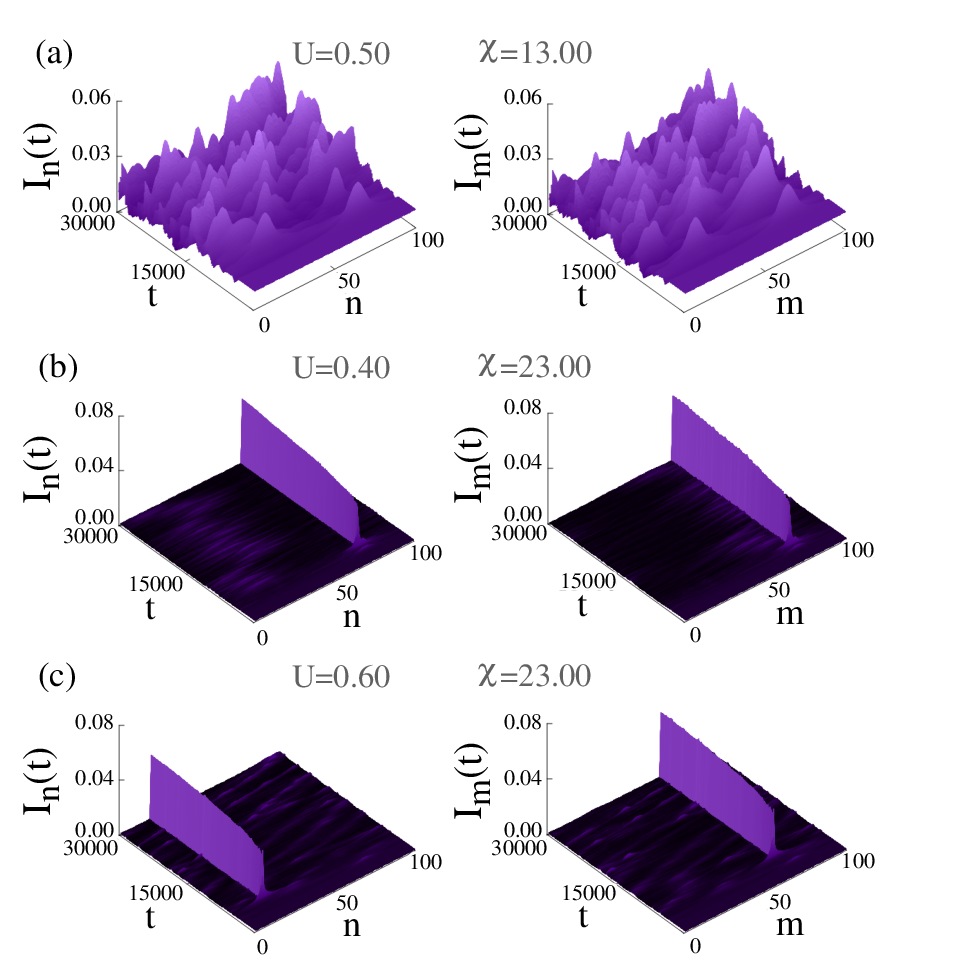}
    \caption{Time evolution of the marginal intensity profiles for representative values of $\chi$ and $U$, illustrating distinct dynamical regimes: (a) Chaotic-like dynamics, characterized by irregular spatiotemporal fluctuations of field across the array; (b) Bound self-trapping, in which the optical field remains spatially localized along the diagonal waveguides after an initial transient; (c) Unbound self-trapping, described by persistent localization of the field at off-diagonal waveguides.}
    \label{fig:5}
\end{figure}

This interplay between bound and unbound states, combined with the non-dispersive character of the nonlinearity, gives rise to additional breathing modes (illustrated in Fig.~\ref {fig:3}g) and yields more intricate dynamical patterns consistent with chaotic-like behavior, as shown in Fig.~\ref{fig:5}a. In this regime, the competition among nonlinearity, evanescent coupling, and Hubbard-like interaction disrupts dynamical regularity, facilitating the excitation of multiple breathing modes and leading to a nonuniform spatial spreading of the optical field. This dynamical instability reflects the system’s sensitivity to perturbations. In addition to the previously reported intermediate regimes, a further analysis also reveals two distinct patterns within the self-trapped regime, both achievable through strong enough nonlinearities. For weak Hubbard-like interactions, self-trapping occurs predominantly along the diagonal waveguides ($n = m$), as shown in Fig.~\ref{fig:5}b. As the interaction strength $U$ increases, self-trapped states emerge in which the optical field becomes concentrated away from the diagonal, indicating an effective suppression of diagonal coupling induced by increased phase mismatch  (see Fig.\ref{fig:5}c).

\begin{figure}[!t]
    \centering
    \includegraphics[height=6.5cm]{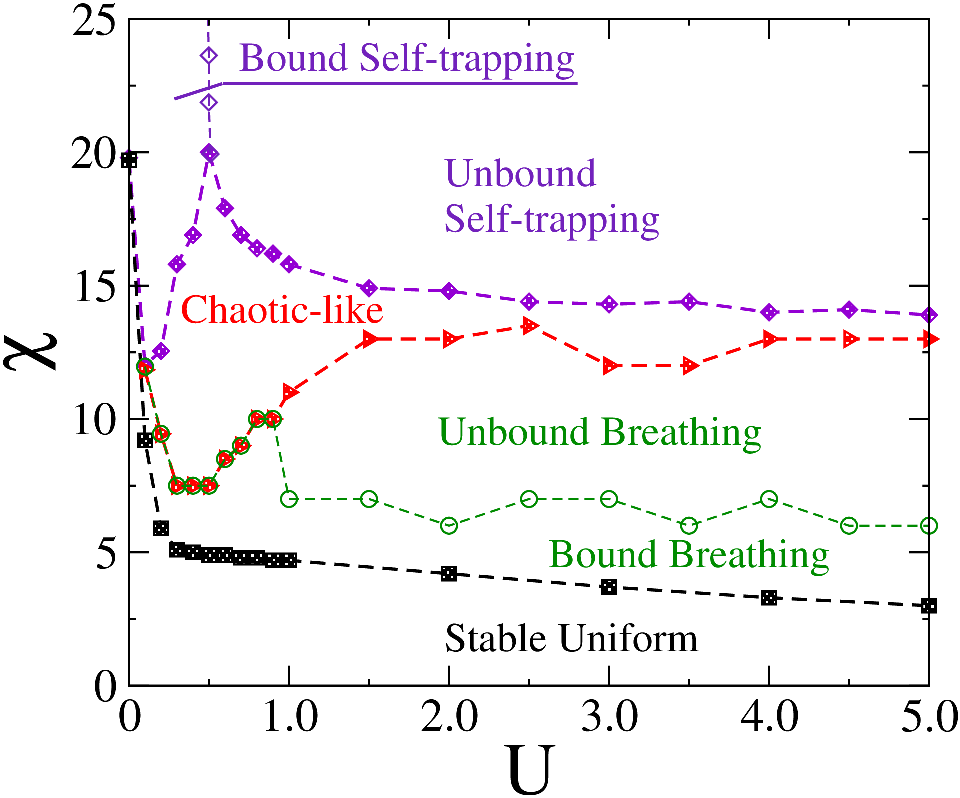}
    \caption{Phase diagram in the $\chi$ versus $U$ parameter space, summarizing the dynamical regimes. The Hubbard-like interaction ($U$) significantly reduces the stability domain of uniform optical distributions. Intermediate values of $\chi$ and $U$ give rise to either well-defined breathing dynamics or chaotic-like behavior, depending on the balance between nonlinearity and interaction. At higher nonlinearities, two distinct self-trapped regimes emerge: one with the field confined along the diagonal waveguides (bound) and another with localization at off-diagonal sites (unbound).}
    \label{fig:6}
\end{figure}

To provide an overview of the possible dynamical behaviors, we show in Fig.~\ref{fig:6} the phase diagram in the $\chi~\mbox{versus}~ U$ parameter space. In the absence of Hubbard-like interactions ($U=0$), the transition from stable uniform states to self-trapped configurations occurs directly, with a threshold at $\chi \approx 19.7$, in agreement with previous studies~\cite{Chaves2015}. When interactions are introduced ($U > 0$), this threshold is significantly reduced, and intermediate breathing regimes emerge, mediating the transition to self-trapping. In general, increasing the nonlinear parameter $\chi$ results in a sequence of distinct dynamical behaviors. For small $\chi$, the system remains in a stable regime, as the nonlinearity is insufficient to overcome the initial delocalization induced by evanescent coupling,  even when considering the penalty associated with Hubbard-like interaction. As $\chi$ increases, it begins to amplify local intensity variations along the diagonal waveguides, thereby supporting bound breathing modes. Further enhancement of $\chi$ disrupts this localized coherence, enabling coupling to off-diagonal sites and giving rise to unbound breathing modes. By continuing to increase the nonlinear parameter, the simultaneous excitation of multiple breathing modes with distinct frequencies disrupts the temporal coherence and gives rise to chaotic-like dynamics. Ultimately, when $\chi$ becomes strong enough, the nonlinear response dominates the dynamics, effectively suppressing the spreading and leading to optical self-trapping characterized by strong spatial localization of the optical field.

We observe a predominance of chaotic-like breathing dynamics in the regime characterized by weak Hubbard-like interactions ($U \lesssim \kappa$). In this scenario, the diagonal waveguides remain sufficiently coupled to the surrounding lattice, enabling substantial interference between diagonal (bound) and off-diagonal (unbound) optical modes. The absence of clear spectral separation between bound and unbound modes suppresses the emergence of well-defined unbound breathing regimes. Thus, increasing nonlinearity drives the system directly from coherent diagonal breathing to irregular, chaotic-like dynamics.  

The competition between bound and unbound states also manifests in the self-trapped regimes. For very weak Hubbard-like interactions, the bound-state subband is embedded within the main band. In this case, the diagonal waveguides remain strongly coupled to the surrounding lattice, enabling efficient exchange between diagonal (bound) and off-diagonal (unbound) modes. However, the slight increase in refractive index along the diagonal, combined with third-order Kerr nonlinearity, establishes a preferential condition for guiding the formation of localized optical mode distributions along the diagonal waveguides (bound self-trapping). As $U$ increases, part of the bound-states subband positions energetically outside the main band. This partial spectral detachment, combined with the effective refractive index barrier imposed by the Hubbard-like interaction, inhibits the formation of localized states and raises the nonlinear threshold for self-trapping. However, beyond $U\approx 0.5$, the mobility blockade imposed by Hubbard-like interaction effectively lowers the number of accessible configurations and promotes self-trapping at lower nonlinearities. In this regime, the optical field localizes predominantly at distinct, off-diagonal sites, characterizing the unbound self-trapping regime.

\section{Summary and concluding remarks}\label{sec:conclusions}

Within a discrete third-order nonlinear Schrödinger framework for finite two-dimensional lattices, we have characterized modulational instability in nonlinear optical waveguide arrays. We focus on the emergent self-trapping regime arising from initially uniform distributions, and demonstrate how engineered diagonal refractive-index detuning, which effectively emulates a photonic analog of on-site Hubbard interactions ($U$) in quantum lattice models, controls this instability threshold and subsequent dynamics pattern. Our results reveal that the presence of such Hubbard-like interactions induces rich dynamical behavior, fundamentally altering the transition between uniform and self-trapped regimes. While systems with $U=0$ undergo a direct transition from uniform to localized (self-trapped) states, the inclusion of Hubbard-like interactions promotes the emergence of intermediate breathing dynamics. These encompass both regular and chaotic-like breathing regimes, driven by the competition between bound and unbound modes. Furthermore, we have identified two distinct self-trapped regimes: a bound self-trapping, where the field localizes along the diagonal waveguides, and an unbound self-trapping, in which the optical field becomes enclosed at off-diagonal sites. A comprehensive phase diagram in the $\chi$ \textit{versus} $U$ space summarizes these regimes, revealing how the Hubbard-like interaction significantly reduces the stability domain of uniform optical distributions while their intricate interplay with Kerr nonlinearity delineates the critical thresholds for each dynamical regime.

These results demonstrate the possibility of controlling nonlinear self-trapped transitions through photonic Hubbard analogs, offering direct implications for optical switching and programmable quantum simulation platforms~\cite{PhysRevLett.106.163901,Karpov2019,Wang2020}. The delineated phase boundaries and stability criteria could guide future experimental implementations in time-multiplexed photonic lattices or integrated optical circuits, particularly in the search of engineered nonequilibrium states and interaction-driven dynamical phenomena.




\section{Acknowledgments}

This work was partially supported by CAPES (Coordena\c{c}\~ao de Aperfei\c{c}oamento de Pessoal do N\'ivel Superior), CNPq (Conselho Nacional de Densenvolvimento Cient\'ifico e Tecnol\'ogico), and FAPEAL (Funda\c{c}\~ao de Apoio \`a Pesquisa do Estado de Alagoas).

%
\section*{Conflict of interest}
The authors declare that they have no known competing financial interests or personal relationships that could have appeared to influence the work reported in this paper.

\bibliographystyle{spphys}
\bibliography{ref}


%
%

\end{document}